# The Effect of Multi-core Communication Architecture on System Performance

Bilal Habib, Ahmed Anber
Electrical & Computer Engineering Department
George Mason University, George Washington University, USA
bhabib@gmu.edu, anbar@gwmail.gwu.edu

Sultan Daud Khan
Computer Engineering Department
Umm Al-Qura University, Saudi Arabia
sgkhan@uqu.edu.sa

*Abstract*— MPSoCs are gaining popularity because of its potential to solve computationally expensive applications. A multi-core processor combines two or more independent cores (normally a CPU) into a single package composed of a single integrated circuit (Chip). However, as the number of components on a single chip and their performance continue to increase, a shift from computation-based to communication-based design becomes mandatory. As a result, the communication architecture plays a major role in the area, performance, and energy consumption of the overall system. In this paper, multiple soft-cores (IPs) such as Micro Blaze in an FPGA is used to study the effect of different connection topologies on the performance of a parallel program.

*Index Terms*— **Multi-core MPI, Performance, Topology**.

## 1. INTRODUCTION

For years, increasing clock speed delivers high performance for wide range of applications. Many applications become more and more complex and require a large amount of computation, single processor cannot frequently satisfy the performance criteria and the designer needs to use multiple processors. A heterogeneous MPSoC consists of two or more independent and distinct microprocessors (cores), i.e., heterogeneous multi-core processors. Eight-core processors are now the norm for desktop workstations and 12-core processors for high-end servers [10]. Some of chip producers, as Tilera® and Plurality, are now manufacturing chips with more cores. Plurality produces a chip with 256 cores. Also, the TILE64™ family of multi-core processors delivers immense computational performance to drive the latest generation of embedded applications. This revolutionary processor features 64 identical processor cores (tiles) interconnected with Tilera's iMesh™ on-chip network. From the software side, it is not surprising that Microsoft's Windows® 7 is 256-core aware operating system and Ubuntu (a Linux based operating system) got an SMP kernel which exploits the cores of the underlying hardware. The challenge
now is to be able to adapt the programming models so that the software makes use of the new operating models emerging from multi-core.However, as the number of components on a single chip and their performance continue to increase, a shift from computation-based to communication-based design becomes mandatory. As a result, the communication architecture plays a major role in the area, performance, and energy consumption of the overall system. we used a multi-core chip that is able to modify its communication architecture. The best technology that can give us this degree of freedom is the FPGAs. The main idea is to fit multiple soft-cores (IPs) in an FPGA and study the effect of different connection topologies on the performance of a parallel program. In this project, we considered using MicroBlaze and Ultrasparc processors and we chose MicroBlaze as it has smaller footprint on the FPGA die so that we can fit multiple Micro Blazes in the same FPGA. That was not the case for using the Ultrasparc processor as one Ultrasparc processor consumed most of the FPGA resources. MicroBlaze processors were interconnected using Fast Simplex Link (FSL) Buses.

The rest of this document is organized as follows: Section 2 discusses related work, Section 3 shows Micro Blaze hardware details. Section 4 describes the different connection topologies and routing mechanisms. This section will also describe the FSL communications. In Section 5 power analysis is described. Device utilization is explained in section 6. And finaly Performance graph in explained 7.

## 2. RELATED WORK

In [1] a multi-core system is tested for parallel algorithms in which the data transfer time is substantially lower than the computing time, and that soft-core processors are appropriate for building multiprocessor systems on a chip. In [3] a cpi application is distributed with MPICH2 libraries achieving a speedup of 3X for using 3 CPUs vs. a single CPU. In this paper, we focus to study the effect of communication latency on the overall system performance. There is also a considerable work done to employ Network on chip (NoC) and emulate it on FPGAs. Some used partial reconfiguration to reduce the emulation time as done in [12]. In [7], the ARM based cores are used for a (NoC) architecture and performance evaluation is carried out for high speed Altera device. In [8], a simulation of NoC is carried out using virtualization. While in [9], a NoC based tool is proposed for architecture exploration. In [11] a functional simulator is proposed that attains the performance benefits of FPGA hardware speed for the common operations (Like, ALU operations), while relegating the infrequent tasks to software. In [13] a time-multiplexing approach has been used to simulate a shared memory multicore systems on FPGA. Which helps in reducing the resources on FPGA in terms of an area.

## 3. HARDWARE ARCHITECTURE

The MicroBlaze core is a 32-bit RISC Harvard architecture soft processor core with 32 general purpose registers, ALU, and a rich instruction set optimized for embedded applications as shown in Fig 1. It supports both on-chip block RAM and/or external memory. The Micro Blaze core implements Harvard architecture. It means that it has separate bus interface units for data and instruction access. Each bus interface unit is further Split into a Local Memory Bus (LMB) and IBM's On-chip Peripheral Bus (OPB). The LMB provides single-cycle access to on-chip dual-port block RAM. The OPB interface provides a connection to both on-and off-chip peripherals and memory. The MicroBlaze core also provides 16 input and 16 output interfaces to Fast Simplex Link (FSL) buses.

*FSL Interface:*

Micro Blaze contains sixteen input and sixteen output FSL interfaces. As shown in Fig 2, the FSL channels are dedicated unidirectional point-to-point data streaming interfaces. The FSL interfaces on Micro Blaze are 32 bits wide.

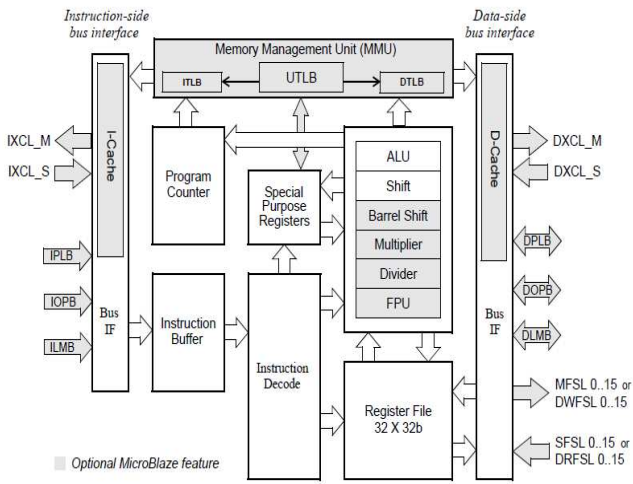

Figure 1: Micro Blaze (v7.10) Core Block Diagram [5]

Furthermore, the same FSL channels can be used to transmit or receive either control or data words. The performance of the FSL interface can reach up to 300 MB/sec [1]. This throughput depends on the target device itself. The FSL bus system is ideal for Micro Blaze-to-Micro Blaze or streaming I/O communications.

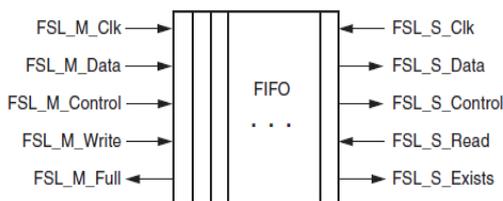

Figure 2: Fast Simplex Link (FSL) Bus Block Diagram [6]

Xilinx EDK provides a set of macros for reading and writing to or from an FSL link. There are two ways of reading/writing on an FSL link: blocking or not blocking, and also there are different instructions for reading/writing data words or control words. The MicroBlaze core is connected to a 64K block RAM (BRAM) which is divided to instruction BRAM and data BRAM. The connection is made to Local Memory Buses (LMB), one for each part of the BRAM. Also all the MicroBlaze cores are connected to a UART controller interface via a Processor Local Bus (PLB). All the cores use the UART for the standard input and output. Each MicroBlaze core also has at least two more FSL connections, to connect it to at least one more core. One FSL on which the core acts as the master and the other on which it acts as slave.

The Micro Blaze architectures consist of eight cores as shown in Fig 2. The only thing that could vary is the number of the FSL links connected to the core as this of course depends on the connection topology and the position of the core in the multi-core network. In Star, a single interrupt controller is used. As shown in Fig 4, only the master node needs to be interrupted in Star topology from different slave nodes.

## 4. COMMUNICATION ARCHITECTURE

There are several types of communication topologies that we studied, such as, Bus, bidirectional ring, star, mesh, torus, and hypercube. We began by implementing three of them and we will further add more later on. The three topologies that are implemented now are cube, ring and star. Fig 5 shows the three connection topologies.

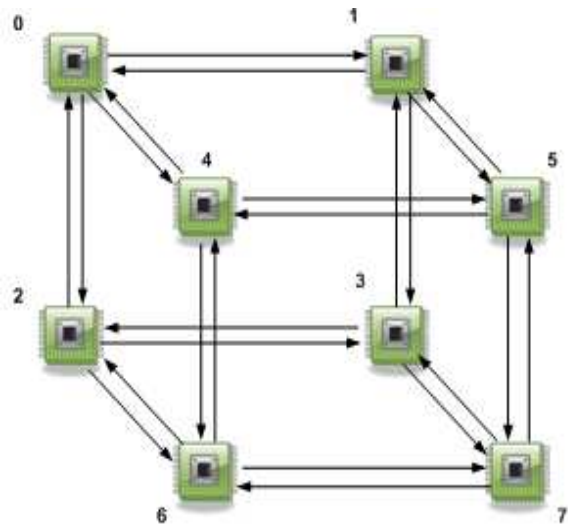

Figure 2 a. Eight-nodes in cube topology with connections.

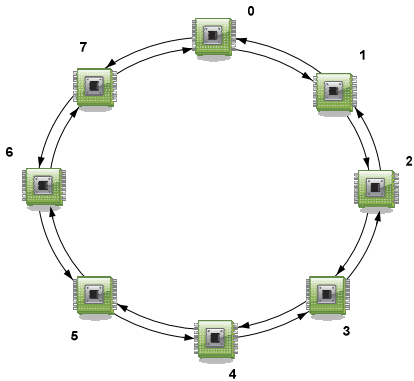

Figure 2 b. Eight-nodes in ring topology with connections.

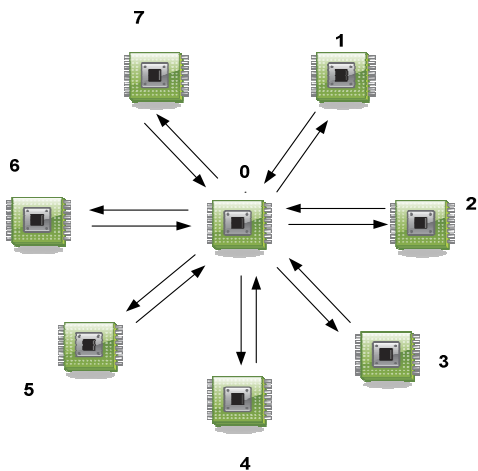

Figure 2 c. Eight-nodes in star topology with connections.

In the Fig 2, each square represents a MicroBlaze with a complete set of peripherals as described in the previous section. There are two FSL link between every two connected cores because FSL is unidirectional. So to support two-way communication, two FSLs have been used, one in each direction. The number of cores in the ring can grow as much as to fill the FPGA. However the maximum number of cores for the star connection is 17. This restriction is due to the fact that the maximum number of FSL link that a single core can handle is 16. So the central node of the star can be connected to no more than eight other cores. The cores on any of the topologies are numbered from zero to N-1 where N is the number of cores such that each core in the network has a static unique id. The ring is numbered such that the ids increase clockwise. The center node in the star is given the id zero and each other node takes unique id. The ids are very important because they make the cores aware of there positions in the network. Cores also make decisions while computing according to their ids. Also while sending and receiving, cores address each other by their respective ids.

The communication architecture is based on the store-and-forward concept. So if two cores are not directly connected and decided to communicate with each other, messages has to go through intermediate cores. Each intermediate core stores the message and then forwards it to the next core until it reaches its destination.

Relying on the store-and-forward routing, we needed to implement a routing library. This routing library acted as a layer between MPI and the underlying network topology as shown in Fig 3. So each topology has a specific routing algorithm. The routing of the messages in the star is trivial. Each core sends the messages to the central core zero and if it is not the destination, the central core will forward it to its destination. In the ring topology, the routing is done via the shortest arc. The direction decision rule depends on the ids of the source and destination cores and also on the number of cores in the ring. The rule is summarized as follows:

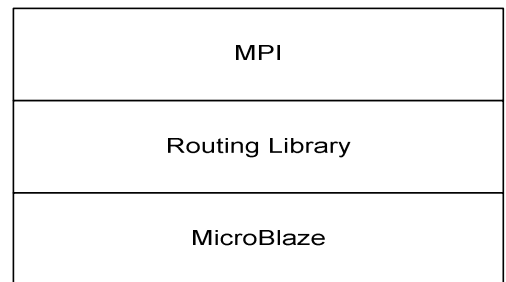

Figure 3. The proposed architecture

The main problem with using the store-and-forward routing is that messages could get stuck in intermediate cores if they were not forwarded them directly. This is because a core could just ignore an incoming message till it finishes its computation. There were two solutions for this problem, one is polling and the other is using interrupts. Polling was rejected for several reasons. First, it is wasting the cores cycles just checking periodically for incoming messages. Second, if there are several intermediate cores, this could severely impact the latency. Third, it is more complex to implement. Thus we used the interrupts to handle incoming messages. So when a message is placed in the inbound FSL of a core, the core is interrupted. The core gets the message from the FSL, if it is not the destination, it will forward it to the next core, and else it will store it in a local buffer to be consumed later.

The problem with interrupts is that MicroBlaze can handle only one interrupt source. So in case, a core is connected to multiple cores, such as the central core in the star topology, MicroBlaze can only sense the interrupt from only one neighbor. The solution to this problem is to use an interrupt controller for each core as shown in Fig 4.

The biggest problem with FSL is that it only handles integer data. To handle real floating point numbers, the number is broken into two parts at the decimal point, and sent as two integers, the value is then assembled at the destination. Before sending the actual number the source core sends two integer values representing the source and destination ids. The destination id will help in routing the message to its designated destination. And the source id will help the destination core recognizes from which core the message is coming from.

When a core receives a message, it checks the destination id, if it is not the destination; the core just forwards the message to the next core according to the routing algorithm. If it is the destination, then it stores the message in an internal buffer and proceeds with its original work.

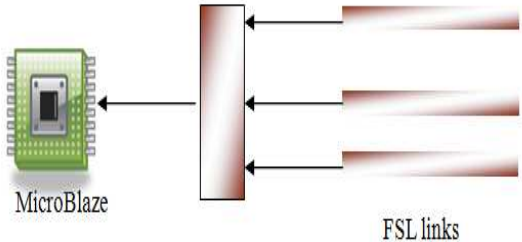

Figure 4. Interrupt Controller enables Micro Blaze to handle more than one interrupt source.

## 5. POWER ANALYSIS

We did power analysis for both topologies at fixed clock frequency of 100Mhz and junction temperature at 54`C. We used Xilinx Xpower Analyzer for power estimation. Power analyses of ring, star and Cube topologies are shown in Table 1 in which all the values are in Watt. Since the Quiescent power is based on the technology which is 65nm in our case, while the dynamic power is based on equation 1 as shown in [4]:

$$\text{Dynamic Power} = CV^2f \quad \ldots(1)$$

According to equation 1, the core frequency is same for all three topologies; therefore the difference in dynamic power is minimal. Which is one of the requirements for fair comparison of three test cases.

|  | TOPOLOGY | | |
| --- | --- | --- | --- |
|  | Ring | Star | Cube |
| Dynamic Power | 1.77018 | 1.76922 | 1.69235 |
| Quiscent Power | 1.26711 | 1.26356 | 1.2633 |
| Total Power | 3.03729 | 3.03278 | 2.95565 |

Table 1: Power consumption

## 6. DEVICE UTILIZATION

We used Xilinx ISE 10.1 to generate the Place and Route reports, which provided us with the device utilization for XC5VLX110t-1ff1136. It is listed for three topologies in Table 2. For fair analysis we employed equal amount of hardware resources, which is evident from the Table 2 figures.

|  | Ring | Star | Cube |
| --- | --- | --- | --- |
| Number of BUFGs | 2(6%) | 2(6%) | 2(6%) |
| Number of DSP48Es | 28(43%) | 28(43%) | 28(43%) |
| Number of External IOBs | 4(1%) | 4(1%) | 4(1%) |
| Number of RAM36 | 128(86%) | 128(86%) | 128(86%) |
| Number of slice Registers | 13630(19%) | 14589(21%) | 13900(20%) |
| Number used as Flip-flops | 13611 | 14572 | 13870 |
| Number of Slice LUTS | 31727(45%) | 30410(43%) | 30572(43%) |
| Number of Slice LUT-Flip flop | 38014(54%) | 37445(54%) | 37690(54%) |

Table 2: Device Utilization

The device utilization figures are primarily based on the number of MicroBlaze cores employed. Since that number is fixed to eight cores in all three topologies therefore the hardware resource consumption closely matches to each other. The number of cores are fixed to eight, because each core has 16 input and output FSL(fast simplex Links), which becomes a bottleneck for Star topology. Hence eight cores are fixed for the remaining two topologies as well.

## 7. PERFORMANCE

We performed the matrix multiplication for different sets of matrix sizes. The performance of star, ring and cube topologies is shown in Fig 5.As shown in Fig 5, the performance of star is better than other two topologies. In our experiment all the computation is done in software i.e in MicroBlaze cores.

For small matrices, all three topologies have almost same performance, but for larger data sets, star outperforms ring and cube. The reason is the excessive communication time that takes place in ring and Cube topologies. Since routing is done in software which helps the Star to broadcast the long matrix size quicker than Ring or Cube. Therefore Star gets the advantage of small communication latency for large data size.

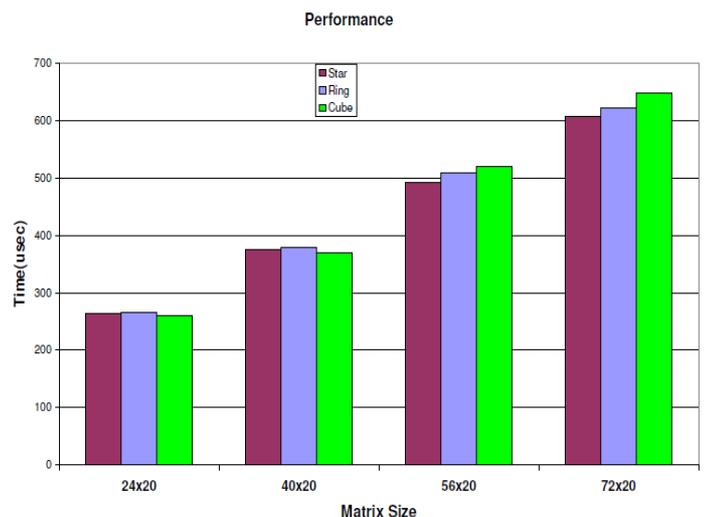

Figure 5: Performance in terms of execution time vs application size

## 8. FUTURE WORK

We need to explore ways to increase the performance of all three topologies and improve the speed-up and efficiency. One way is to do the hardware routing, instead of software routing. Since each send/receive command is operated on a single byte, therefore hardware routing can offer huge improvements.

We also need to analyze more topologies, and their comparison for different applications.